\documentclass[oneside]{article}
\usepackage{graphicx}
\usepackage[utf8]{inputenc}
\usepackage{amsmath, bbm}
\usepackage{amsfonts, amssymb}
\usepackage{amsthm}
\usepackage{xcolor}
\usepackage{graphicx}
\usepackage{caption}
\usepackage{diagbox}
\usepackage{multicol}
\usepackage{enumitem}
\usepackage{authblk}
\usepackage{url}
\usepackage[numbers,square]{natbib}

\usepackage{dcolumn}
\newcolumntype{d}[1]{D{.}{.}{#1}}

\usepackage{lineno}
\modulolinenumbers[5]

\RequirePackage[colorlinks=true,urlcolor=blue!50!black, citecolor=blue!50!black,linkcolor=blue!50!black]{hyperref}

\usepackage{verbatim}

\theoremstyle{plain}
\newtheorem{defn}{Definition}

\theoremstyle{remark}
\newtheorem{rem}{Remark}

\usepackage[textwidth=2.2cm,textsize=footnotesize]{todonotes}

\begin{document}

\title{Probabilistic Models of Profiles for Voting by Evaluation}



\author[1]{Antoine Rolland} 
\author[2]{Jean-Baptiste Aubin}
\author[3]{Irène Gannaz}
\author[2]{Samuela Leoni}

\affil[1]{ERIC EA 3083, Universit\'e de Lyon, Université Lumi\`ere Lyon 2, \protect\\ 5 Pierre Mendès France, 69596 Bron Cedex, France}
\affil[2]{Univ Lyon, INSA Lyon, UJM, UCBL, ECL, ICJ, UMR5208, \protect\\ 69621 Villeurbanne, France}
\affil[3]{Univ. Grenoble Alpes, CNRS, Grenoble INP\footnote{Institute of Engineering Univ. Grenoble Alpes}, G-SCOP,\protect\\ 38000 Grenoble, France}

\maketitle

\begin{abstract}
Considering voting rules based on evaluation inputs rather than preference rankings modifies the paradigm of probabilistic studies of voting procedures. This article proposes several simulation models for generating evaluation-based voting inputs. These models can cope with dependent and non identical marginal distributions of the evaluations received by the candidates. A last part is devoted to fitting these models to real data sets.
\end{abstract}

\section{Introduction }

\label{sec:objectifs}

Voting is a very general method to turn individual preferences on a finite set of candidates into the choice of the best candidate, i.e. the winner of the election. These individual preferences can be expressed as rankings or evaluations on candidates. 
A review of both ranking-based and evaluation-based voting rules can be found in \cite{durand22}. 
Both social and mathematical approaches consider positive or negative properties satisfied by a given voting rule as a matter of interest, such as independence to irrelevant alternatives, Condorcet paradox or manipulability.
  Studying the properties of voting rules can be done either through an axiomatic approach or a probabilistic approach. The axiomatic approach supposes to determine which properties characterize a specific voting rule, \emph{i.e.} which properties are to be observed, and which are not, via formal theorems. The probabilistic approach aims at determining whether these properties are likely to be observed, \emph{i.e.} determining the frequency of occurrence of such properties using a given voting rule.

 Many works have been published on the probabilistic approach. For instance, without claiming to be exhaustive,  \cite{Tideman2012},  \cite{Tideman2014}, \cite{Plassman2014}, \cite{Green-Armytage2016}   contain examples of simulation-based studies of voting rules. Analogously, \cite{gehrlein2006condorcet}
 looks at the frequency of occurrence of Condorcet's paradox. \cite{durand22} studies the manipulability rate of voting systems. \cite{Fav} do the same focusing on coalitional manipulability. Most of these articles consider only voting rules based on rankings on candidates.  
 
One can refer to \cite{Diss2020} for a recent state-of-the-art of generative models for a probabilistic approach of voting theory.  Generative models for ranking-based voting preferences (i.e. cardinal models) can be divided into two main approaches. The first approach consists in directly generating random rankings through a specific distribution on the space of ranking. It is difficult, if not impossible, to generalize existing methods to evaluations because of the very different nature of the information contained in a ranking and in evaluations. The second approach entails working with utilities, incorporating for instance spatial models. As observed by \cite{Critchlow}, one can generate any distribution on rankings from continuous random variables. The \textit{Thurstone order statistics models} (see e.g. \cite[chapter 8]{Alvo} and references therein) aim at building probability distributions on rankings by ordering continuous random variables, named utilities. 

In continuous evaluation-based models, the grades can be viewed as bounded utilities. Similarly to Thurstone models, a strict preference relation on the candidates can be obtained from the evaluations for each voter. 
In case of discrete grades, \emph{ex-aequo} may occur, and then a pre-order can be obtained from evaluations, but not a complete order. The literature is mainly focused on simulations for ranking-based voting rules, while our main matter of interest is evaluation-based voting rules. 
Evaluation-based voting rules, including the three main ones: range voting, see \cite{Smith00rangevoting}, majority judgment, see \cite{Balinski2007}, and approval voting, see \cite{Brams2007}, but also the new family of "deepest voting", see \cite{AUBIN2022}, are based on the use of  evaluations on a bounded scale (which can be discrete or continuous) given by each voter to each candidate.  Generative models for evaluation-based voting rules did not get all the attention they deserve in our opinion : filling this lack is our main motivation in this paper.  
 
 Our objective here is therefore to propose generative models for evaluation-based voting rules, in the  framework of a probabilistic approach.  We model voting situations through the rationalization of probabilistic distributions of evaluations upon candidates.
We first propose in Section~\ref{sec:IC} some distributions for univariate evaluations for one candidate.
A social interpretation of the univariate evaluations distributions will be given in Section~\ref{sec:interpret}. Next, Section~\ref{sec:multi} is devoted to the distributions for multivariate evaluations. The simplest approach is to consider independent and identically distributed (i.i.d.) voters' evaluations on each candidate. This assumption can be relaxed to enlarge the scope of generating models. To this aim, we introduce in Section~\ref{sec:dep} copula-based evaluations models and spatial models. Section~\ref{sec:geneobs} provides an illustration on fitting on observed situations. 

\subsection*{Notations} 

In the following, we will consider situations with $n$ voters and $m$ candidates. Each voter associates an evaluation from a set $\mathcal E$ to each candidate.

Evaluation of voter $v$ for candidate $c$ will be denoted $e_{vc}$, for $v=1,\dots,n$, $c=1,\dots,m$. Observations $\{e_{vc}, ~{v=1,\dots,n ,~ c=1,\dots,m}\}$ are $n$ independent realizations of a random variable $E=(E_1,\ldots,E_m)$, which takes values in $\mathcal E^m$. 
Note that we will not consider dependence among the voters, as discussed later, in Section \ref{sec:dep_voters}.

Under the assumption of independence between voters, defining a simulation setting can be seen as defining a multivariate probability distribution on $\mathcal E^m$. 
We will consider two cases with respect to the amount of information contained in the set $\mathcal E$: 
\begin{itemize}
    \item continuous grades: without loss of generality, $\mathcal E=[0,1]$ ;
    \item discrete grades: without loss of generality, $\mathcal E=\{0,\dots , K\}$ with $K \in \mathbb{N} \setminus \{ 0 \} $. 
\end{itemize}

The easiest distributions one can consider for the $m$ random variables $E_1,\dots E_m$  are independent and identically distributed (i.i.d.) distributions. When the evaluations are continuous, the preferences resulting from the evaluations will satisfy almost surely an Impartial Culture (IC)  model, that is, a uniform distribution on rankings. 

This model can be extended to more complex ones following two ways. The first way is to remove the identical distribution assumption: two candidates may not have the same evaluation distribution in the population of voters. The second way is to introduce dependence between the evaluations of the candidates. Indeed, two candidates viewed as almost similar (resp. in opposition) by the voters may lead to positively (resp. negatively) correlated evaluations.

\section{Univariate distributions on evaluations} \label{sec:IC}

We propose in this section several distributions to model the evaluation distribution on a single candidate by several independent voters, respectively for continuous and discrete sets. A social interpretation of these models is provided in Section~\ref{sec:interpret}. Independence between voters is justified in Section \ref{sec:dep_voters}.
 
\subsection{Continuous case}

\begin{figure}[!ht]
    \centering
\includegraphics[width=0.24\textwidth]{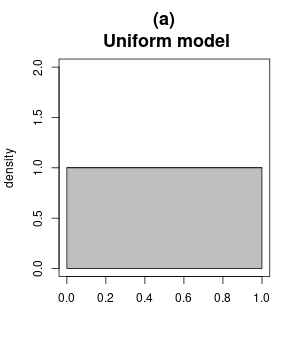}
    \includegraphics[width=0.24\textwidth]{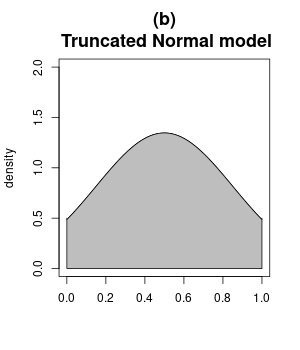}
    \includegraphics[width=0.24\textwidth]{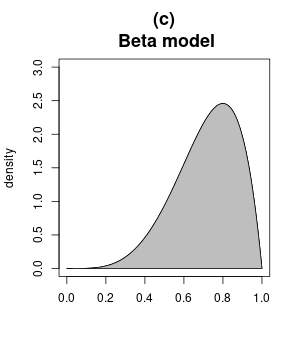}
    \includegraphics[width=0.24\textwidth]{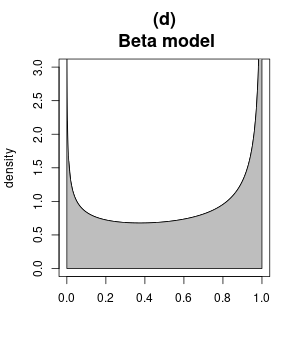}
    \caption{Probability distribution functions of evaluations on one candidate. (a) Uniform distribution, (b) Truncated Normal distribution with $\mu=0.5$ and $\sigma=0.35$, (c) Beta distribution with $\alpha=5$ and $\beta=2$ and (d) Beta distribution with $\alpha=0.7$ and $\beta=0.5$.}
    \label{fig:exempleEvIC}
\end{figure}

When the evaluations are continuous, that is, $\mathcal E=[0,1]$, we propose to use Uniform, truncated Normal or Beta distributions, as illustrated in Figure~\ref{fig:exempleEvIC}. 
Uniform distribution is the easiest model. Truncated Normal distribution is useful to model uni-modal distributions when the proportions of evaluations close to 0 or 1 are significant. Beta distribution is a very versatile model that can model several situations, depending on the chosen parameters; it is particularly adapted in the case of bi-modal distributions, when most of the evaluations are around 0 or 1, but also in the case of uni-modal distribution with a quick decreasing curve. 
 
It can be seen that these three families of distributions cover a large scope of distributions.
The initial distribution proposed by \cite{Thurstone} is the standard Gaussian distribution. Our truncated Normal distribution is, hence, a generalization to bounded-support evaluations. Other distributions were introduced in Thurstone's like approach. The most famous is the Gumbel distribution proposed by \cite{Luce}, which has the advantage of providing a closed form distribution of rankings. Yet, Gumbel distribution does not have a bounded support and, hence, cannot be used in our context.

\subsection{Discrete case}

\begin{figure}[!ht]
    \centering
\includegraphics[width=0.24\textwidth]{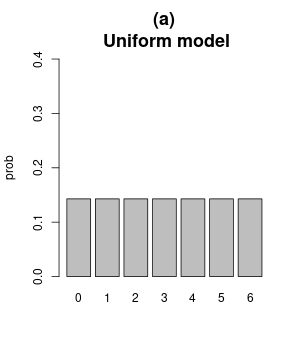}
    \includegraphics[width=0.24\textwidth]{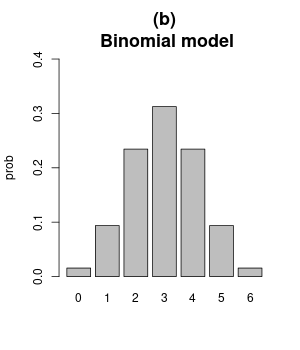}
    \includegraphics[width=0.24\textwidth]{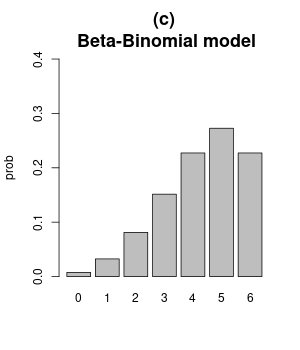}
    \includegraphics[width=0.24\textwidth]{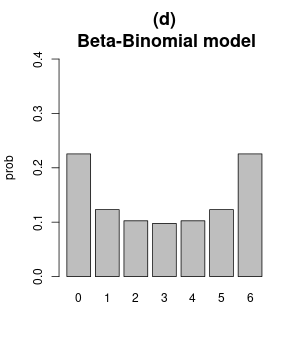}

\caption{Examples of probability distributions on 7 levels ($K=6$). (a) Uniform distribution, (b) Binomial distribution with $p=0.5$, (c) Beta-Binomial distribution with $\alpha=5$ and $\beta=2$, and (d) Beta-Binomial distribution with $\alpha=\beta=0.5$.}
    \label{fig:exempleEvICdiscrete}
\end{figure}

When the evaluations are discrete, i.e. $\mathcal E=\{0,\dots,K\}$, we propose to use discrete Uniform, Binomial or Beta-Binomial distributions, which can be seen as the discrete counterpart of the Uniform, truncated Normal and Beta continuous distributions.
Examples are presented in Figure~\ref{fig:exempleEvICdiscrete}.

\subsection{General remarks}
\label{sec:interpret}
Parameters for each model are summed up in Table \ref{tab:distrib}.

\begin{table}[!ht]
\begin{tabular}{lc}
Marginal continuous distributions &  Parameters \\ \hline
    Continuous Uniform $\mathcal U[0,1]$ &  $\cdot$ \\
         Truncated Normal $\mathcal N_T(\mu,\sigma^2)$  & $(\mu, \sigma)$  \\
    Beta $\mathcal B(\alpha,\beta)$ &  $(\alpha,\beta)$  \\
\end{tabular}
\hfill
\begin{tabular}{lc}
Marginal discrete distributions &  Parameters \\ \hline
Discrete Uniform $\mathcal U\{0,\dots,K\}$ &  $\cdot$   \\
        Binomial $\mathcal B(K,p)$ &  $p$ \\
        Beta-Binomial $\mathcal B(K,\alpha,\beta)$ &   $(\alpha,\beta)$ \\ 
\end{tabular}
\caption{Marginal distributions proposed for the distribution of the evaluations. $K$ is not a flexible parameter since it corresponds to the known scale of the evaluations.}
\label{tab:distrib}
\end{table}

In a similar way to traditional statistical studies, the distribution of evaluations given to candidates can be chosen according to two main parameters: a location parameter and a dispersion parameter.
The location parameter (e.g. the mean or median of evaluations) will be proportional to a candidate's general popularity. For example, a candidate whose evaluations distribution is that of Figures~\ref{fig:exempleEvIC}(c) or \ref{fig:exempleEvICdiscrete}(c) will be more popular than a candidate whose distribution is that of case Figures~\ref{fig:exempleEvIC}(b) or \ref{fig:exempleEvICdiscrete}(b).
The dispersion parameter indicates a candidate's ability to divide opinions. In the extreme case of a zero dispersion, all voters would associate the same score to the candidate. Conversely, a high dispersion, like in Figures~\ref{fig:exempleEvIC}(d) or \ref{fig:exempleEvICdiscrete}(d), is associated with a candidate that divides opinion.
It is therefore possible to associate one distribution rather than another {\em a priori}, depending on the information we have on the candidates, with the uniform distribution always being preferred if we have no information on the candidate's evaluations.

Of course, these distributions are not exhaustive and many choices are possible. 
We focus here on the most natural, which already seem to cover various shapes. An extension can be proposed by the use of a mixture model of distributions. For example, \cite{Murphy} proposed mixture models for rankings. This is especially interesting when fitting models on real voting elections, see \cite{Dubin} or \cite{Gormley}. As our main objective is to propose generating models, we do not explore this direction here. But note that none of the distributions presented here is multimodal without modes in the extremities. This is a limit for the fitting of real data, that mixtures can overcome.

The earlier methods require certain assumptions about the distribution of the evaluations, which must be validated. These assumptions can be assessed using goodness-of-fit tests such as the Chi-Square test, Kolmogorov-Smirnov test, Anderson test, and others. In instances where the theoretical distribution does not align with the observed data, more versatile techniques become pertinent. Specifically, non-parametric approaches like kernel density estimation or projection-based estimation of evaluation densities can be employed. This spectrum of methods ensures both adaptability and verifiability of the approach.

\subsection{Dependence between voters}
\label{sec:dep_voters}

Some simulation approaches are based on introducing dependencies between the voters, see e.g.  \cite{dep_voters_berg,dep_voters_ladha,dep_voters_berend,dep_voters_chollete}. This corresponds to the so-called \textit{multi-stage} approach, see \cite[Chapter 8]{Alvo}. This approach enables to generate non Uniform distribution on rankings. It can be used for example to obtain Mallow's Phi distribution. Its main interest is the ease of the simulation process. 
The objective of such models, rather than controlling the dependence between the voters, is to obtain a given distribution on the rankings.

\begin{figure}
\includegraphics[width=\textwidth, height=5cm]{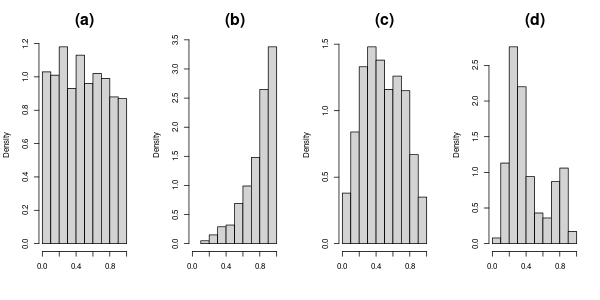}
\caption{Influence of the dependence between voters on evaluation distributions. Evaluations are defined by $e_{v}=\Phi(x_v)$ with $(x_1,\dots,x_m)\sim\mathcal N(0,\Sigma)$, with $m=1000$, and $\Phi$ the cumulative distribution function of the standard Gaussian distribution. In subfigures (a) to (c), for all $(v,v')\in\{1,\dots,m\}^2$, $v\neq v'$, $\Sigma_{v,v}=1$, $\Sigma_{v,v'}=\rho$. From left to right, (a) $\rho = 0$, (b) $\rho = 0.5$, (c) $\rho = 0.5$ with a different seed in simulation. In subfigure (d), $\Sigma$ is a block matrix, corresponding to 2 communities : $\Sigma_{v,v}=1$ for all $v=1,\dots,1000$, and if $v\neq v'$, $\Sigma_{v,v'}=0.9$ if $1\leq v,v'\leq 250$ or $251\leq v,v'\leq 1000$, and $\Sigma_{v,v'}=0$ else.}
\label{fig:dep_voters}
\end{figure}

Introducing the dependencies between the voters does not bring relevant information for the generation of evaluations. The dependence between voters yields a different distribution of the evaluations, and this information is lost when considering only the resulting values since the voters are considered anonymous. An example is displayed in Figure~\ref{fig:dep_voters}. Hence, from a generating point of view, it is more interesting to model the structure of the voters through the distribution (with mixing of distributions to model communities - see Figure~\ref{fig:dep_voters}(d)). From a fitting point of view, it is not possible to estimate dependence between voters. Indeed, estimating a dependence needs to have repeated observations of a variable. 

\begin{rem} Evaluations $\lbrace{e_{v}\rbrace}$ in Figure~\ref{fig:dep_voters} are dependent observations from a Uniform distribution on [0,1]. Indeed, if a random variable $X$ follows a standard Gaussian distribution, and $\Phi$ is its cumulative distribution function, than $\Phi(X)$ follows a Uniform distribution on $[0,1]$. Repeating simulations with different seeds, the resulting distribution would indeed be uniform. As shown in the figure, if only one sample is observed, retrieving the uniform distribution (as well as the dependence structure) is impossible. 
\end{rem}

\section{Multivariate evaluation models}\label{sec:multi}
\label{sec:multivariate}

Generating evaluations on several candidates for each voter need to define a multivariate distribution of evaluations. This section introduces distributions for the random variables $E_1,\dots, E_m$ corresponding to voters' evaluations for each of the $m$ candidates. First, i.i.d distributions can be considered. Subsequently, the assumption of identical distributions across candidates and the assumption of independence can be progressively relaxed.

The simplest case of  evaluation-based voting process is to consider that voters' evaluations of each candidate are i.i.d., which we will call an Independent and Identically Distributed (IID) evaluations model. 

\begin{defn}{\bf IID  model}~\\ \label{def:EvIC}
 The Independent Identically Distributed (IID) evaluations model is such that random variables $E_1,\ldots, E_m$ are independent and identically distributed. 
\end{defn}
As far as the evaluations are continuous and i.i.d., IID models almost surely yield IC models on rankings. We propose hereafter alternatives to IID models. The first possibility is to deal with independent and not identically distributed distributions. Such models will be called Independent and Differently Distributed (IDD) evaluations Models.


 \begin{defn}{\bf IDD  model}~\\ \label{def:EvIPC}
 The Independent and Differently Distributed (IDD) evaluations model is such that the  random variables $E_1,\ldots, E_m$ are independent and non identically distributed.
\end{defn}

The marginal distributions of $(E_1,\ldots, E_m)$ can be different by changing parameters of a given distribution family, or by changing the distribution family. 
As noticed previously, Beta (continuous) distributions and Beta-Binomial (discrete) distributions enable to consider various shapes of distributions.

A possibility to extend IID models and IDD models is to remove the independence hypothesis between the evaluations of each candidate.
In such a case, the marginal distributions can be either identical or different. 
Hence, these models can be considered as Dependent and Identically Distributed evaluations models (DID). 
\begin{defn}{\bf DID  model}~\\ \label{def:EvDIC}
 The Dependent and Identically Distributed (DID) evaluations model is such that random variables $E_1,\ldots, E_m$ are dependent and identically distributed. 
\end{defn}

Considering non identical distributions provide Dependent and Differently Distributed evaluations models (DDD). 

 \begin{defn}{\bf DDD  model}~\\ \label{def:EvDPC}
 The Dependent and Differently Distributed (DDD) evaluations model is such that the random variables $E_1,\ldots, E_m$ are dependent and non identically distributed. 
\end{defn}

Many DDD models can be proposed. Some are detailed in the next section.

\section{Construction of dependent models}

\label{sec:dep}

Several approaches can lead to a DDD model. 
We propose first to use copula-based models to introduce dependence between candidates. This approach enables to define separately the univariate distributions of each candidate. Next, we introduce multinomial and Dirichlet models to describe the special case of cumulative voting. Finally, spatial models offer a large scope of multivariate distributions.

\subsection{Copula-based models}
\label{sec:copules}

Copulas are useful tools to represent dependencies between variables. If the distributions of the evaluations of each candidate are not independent, a multivariate copula can be used to take into account their dependencies. In a nutshell, a copula is a multivariate cumulative distribution function which has all its margins uniformly distributed on the unit interval. It can also be applied on transform of random variables to generate dependence with non uniform marginals. See \cite{copula-everything,nelsen2006} for a formal presentation of the subject.

\begin{defn}{\bf Copula DDD models}~\\ \label{def:copula}
The Copula DDD models are defined by $$E=(E_1,\ldots,E_m)\sim C(\mathcal{D}_1, \ldots \mathcal{D}_m),$$
where  $C$ is a multivariate copula and $\mathcal{D}_1,\ldots, \mathcal{D}_m$ are distributions on $\mathcal E$.
That is, $C$ is a function from $[0,1]^m$ to $[0,1]$ such that the cumulative distribution function (c.d.f.) of $E$ writes as $C(\Delta_1(E_1),\dots,\Delta_m(E_m))$ with $\Delta_c$ c.d.f. of marginal distributions $\mathcal{D}_c$ of $E_c$, for $c=1,\dots,m$.
\end{defn}

A strength of copulas is that they allow any marginal distributions. Therefore, the model should specify both the marginal distribution for each candidate, and the copula used to model the dependencies between variables. In this sense, Copula DDD models include also IID, IDD and DID models, which appear as particular cases of copula DDD models (see below).

Note that in Thurstone order statistics modeling, with continuous variables, copulas were introduced by \cite{MacFadden} and studied e.g. by \cite{Joe}. In the following, we distinguish the cases of a continuous evaluations set and a discrete evaluations set $\mathcal E$.

\subsubsection*{Choice of a copula in a continuous case}

In the continuous case, 
 Gaussian copulas offer a simple way to model dependencies between each pairs of candidates, through the correlation coefficients. For a given correlation matrix $R\in\mathcal{R}^{m\times m}$, the associated Gaussian copula writes as $ C_R(\mathcal{D}_1, \ldots \mathcal{D}_m)= \Phi_{R}(\Phi^{-1}(\mathcal{D}_1),\dots, \Phi^{-1}(\mathcal{D}_m)),$
where $\Phi^{-1}$ is the inverse c.d.f. of a real standard normal and $\Phi_{R}$ is the joint c.d.f. of a multivariate centered normal distribution with covariance matrix $R$. The dependencies of the copula  between two variables are exactly characterized by the correlation coefficients. This model has therefore the advantage to be easy to simulate and to enable to define in a very comprehensive way the dependence between the evaluations. Another usual family of copulas for the continuous case is Archimedean copulas. We decide not to detail this family of copulas for the sake of simplicity.

Another interesting copula class is the checkerboard copula class, see \cite{Cuberos2020}, which represents a good compromise between the richness of the expression and the complexity of the model. This copula is based on a partition of $[0,1]^m$, $(B_b)_{b\in\mathcal B}$, $\mathcal B\subset \{0,\dots, B-1\}^m$, $B\in\mathbb N\setminus\{0\}$, with $B_b=(b_1/B,(b_1+1)/B)\times\dots\times (b_m/B,(b_m+1)/B)$. On each $B_b$, $b\in\mathcal B$, the copula is constant, equal to the value of the empirical copula at the left corner, that is, the c.d.f. $C(x_1,\dots, x_m)$ is proportional to the cardinal of $\lbrace j,\; (e_{1j},e_{2j},\dots,e_{mj})\in B_b, \; \forall c\in{1,\dots,m}, e_{cj}\leq b_j/B \rbrace $. This copula is, hence, a piecewise constant version of the empirical copula. It reduces the overfitting compared with the empirical copula. We refer \emph{e.g.} to \cite{copula-choice}  for a discussion on the choice of a copula.
    
\subsubsection*{Choice of the copula in a discrete case}

Evaluations on discrete scales need the use of specific discrete copulas for simulation. The discrete copula function $C$ has to be defined only on a finite subset of $[0,1]^m$, which is in our context $\lbrace 0, \frac{1}{K},\dots, \frac{K-1}{K}, 1 \rbrace^m$. Among others, Vines pair-copulas, see \cite{Panagiotelis2012}, and Gaussian copulas, see \cite{Barbiero2017}, have been proposed to simulate dependent discrete data, and therefore can be used also to model discrete evaluations in a social choice framework. In particular, the Gaussian copula is defined as in the continuous case.

\subsubsection*{Marginal distributions}

Additionally, marginal distributions must be chosen. One can consider the same distribution for each candidate and obtain DID models, or different distributions following DDD models. 
Except for the Copula DID Uniform continuous models, which belong almost surely to the IC model since all candidates have the same distributions, each of these models allows for different marginal distributions for the evaluations of each candidate. Additionally to this non identical setting, the dependence modeling, through the copula, yields a large scope of models. These copula-based models appear very rich and adapted for covering much framework of simulations of evaluations.

\begin{figure}[!ht]
    \centering
    \includegraphics[width=0.3\textwidth]{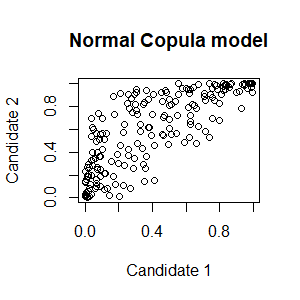}
        \includegraphics[width=0.3\textwidth]{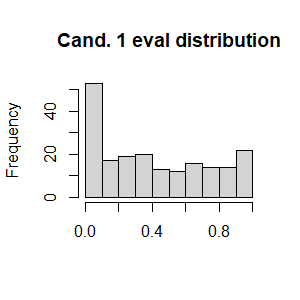}
    \includegraphics[width=0.3\textwidth]{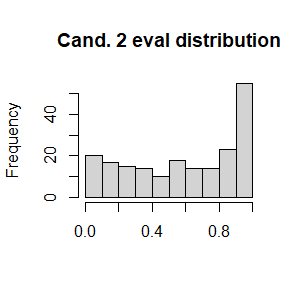}
        \caption{Simulation examples of DDD Copula evaluations for $m=2$ candidates and $n=200$ voters. }
    \label{fig:copula}
\end{figure}

Figure~\ref{fig:copula} presents a simulation based on the use of two  different marginal distributions for two candidates (Beta distribution with parameters (0.7,0.5) for the first candidate and (0.5,0.7) for the second one), using a Gaussian copula of parameter 0.8.

\subsection{Multinomial and Dirichlet models for cumulative voting}

DDD models suppose that there are dependencies between evaluations given by a voter to the candidates. Among these models, a specific case is related to cumulative voting, which consists in dividing a total score on the candidate evaluations, such that the evaluations' sum on the set of candidates is the same for each voter. We refer for example to \cite{cumulative1, cumulative2} and references therein.
In such a context, the evaluations are not independent as there is a link between the evaluations for all the candidates given by a single voter.

Both discrete and continuous models are available, using multinomial distribution (a multivariate version of the Binomial distribution) in the discrete case and Dirichlet distribution  (a multivariate generalization of the Beta distribution) for the continuous case.

When the evaluations are discrete, i.e. $\mathcal E=\{0,\dots,K\}$, the DDD multinomial model is the following.
 \begin{defn}{\bf DDD  multinomial model}~\\ \label{def:multinomial}
 The DDD  multinomial model is defined by
 $(E_1, \ldots , E_m) \sim \mathcal{M}\{K,p_1, \ldots, p_m\}$, where $\mathcal{M}\{K,p_1, \ldots, p_m\}$ is  the multinomial distribution of parameters $K$ and $p_1,\ldots , p_m$, with for all $c=1,\ldots,m$, $p_c \geq 0$ and $\sum_{c=1}^m p_c=1$.
 \end{defn}

The marginal distributions of a multinomial distribution are Binomial distributions; the higher $p_c$, the higher the mean score of candidate $c$, $c=1,\dots, m$. The multinomial model introduces dependence between the evaluations of the candidates. More precisely, if parameters $(p_c)_{c=1,\dots,m}$ do not have the same value, then the evaluations are not identically distributed and the associated multinomial model is then an DDD model. If for all $c =1, \ldots, m$, $p_c=1/m$, then the associated multinomial model is an DID model. Note that $K$ is the maximal authorized grade.

\begin{figure}[!ht]
    \centering
    \includegraphics[width=0.3\textwidth]{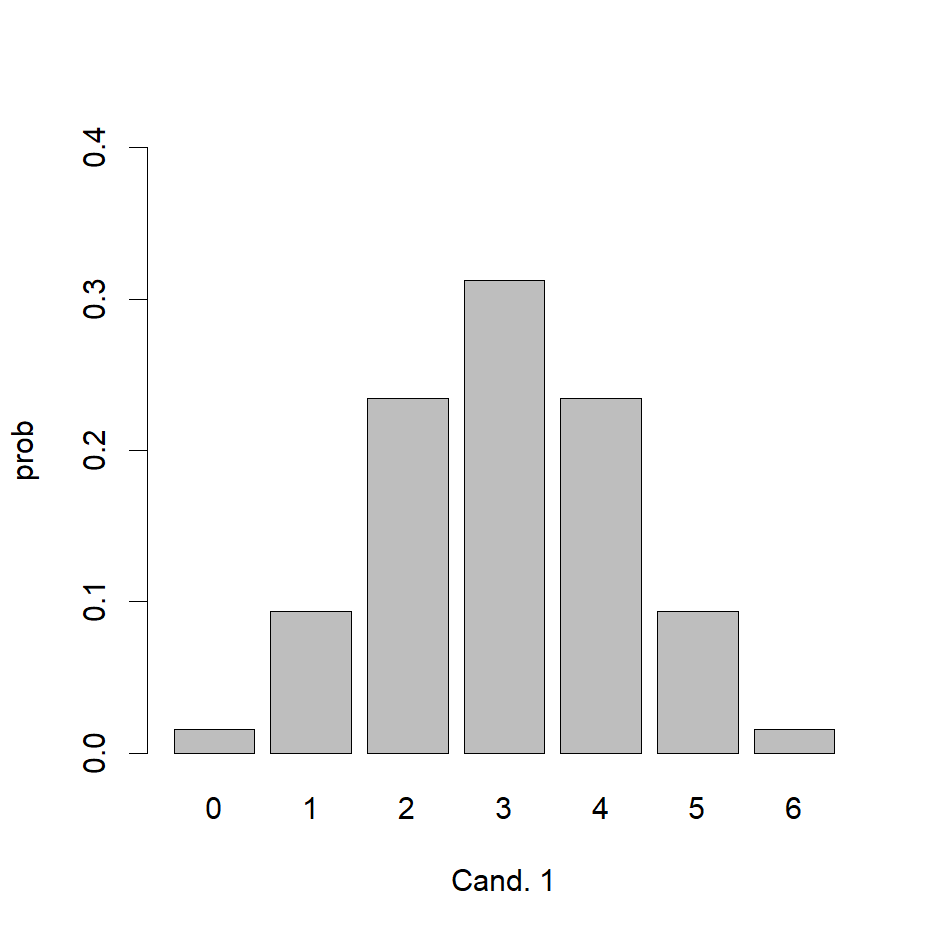}
    \includegraphics[width=0.3\textwidth]{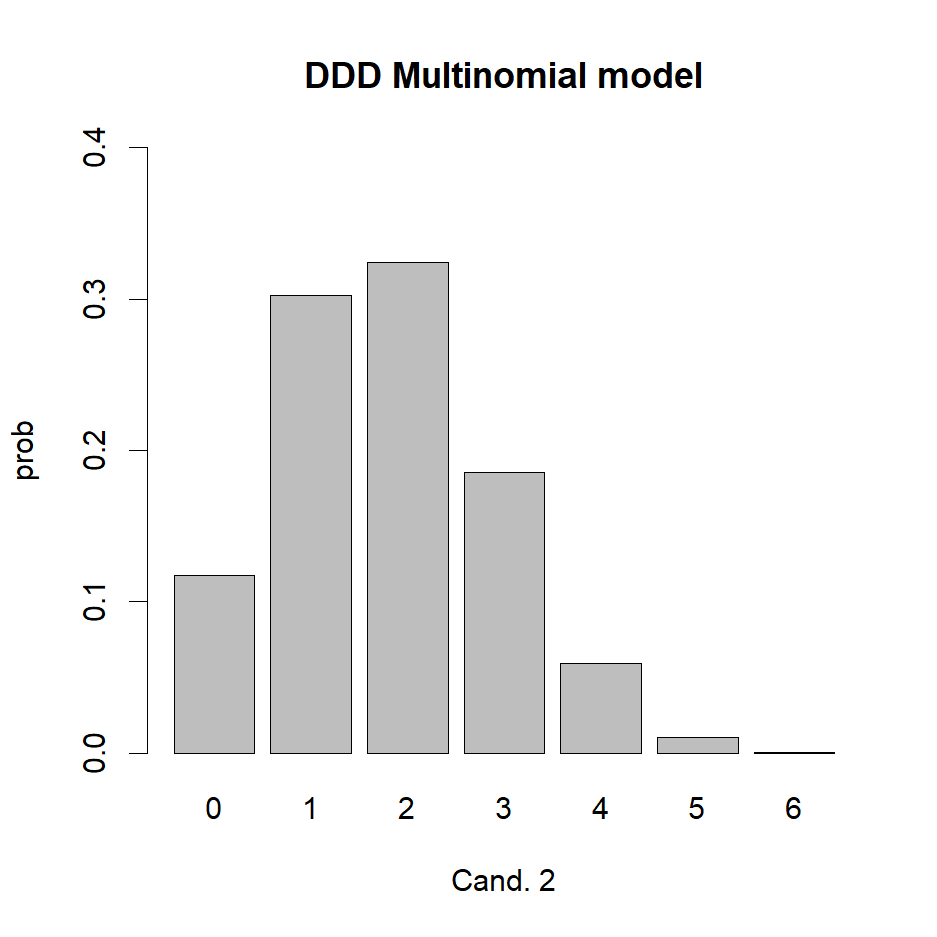}    \includegraphics[width=0.3\textwidth]{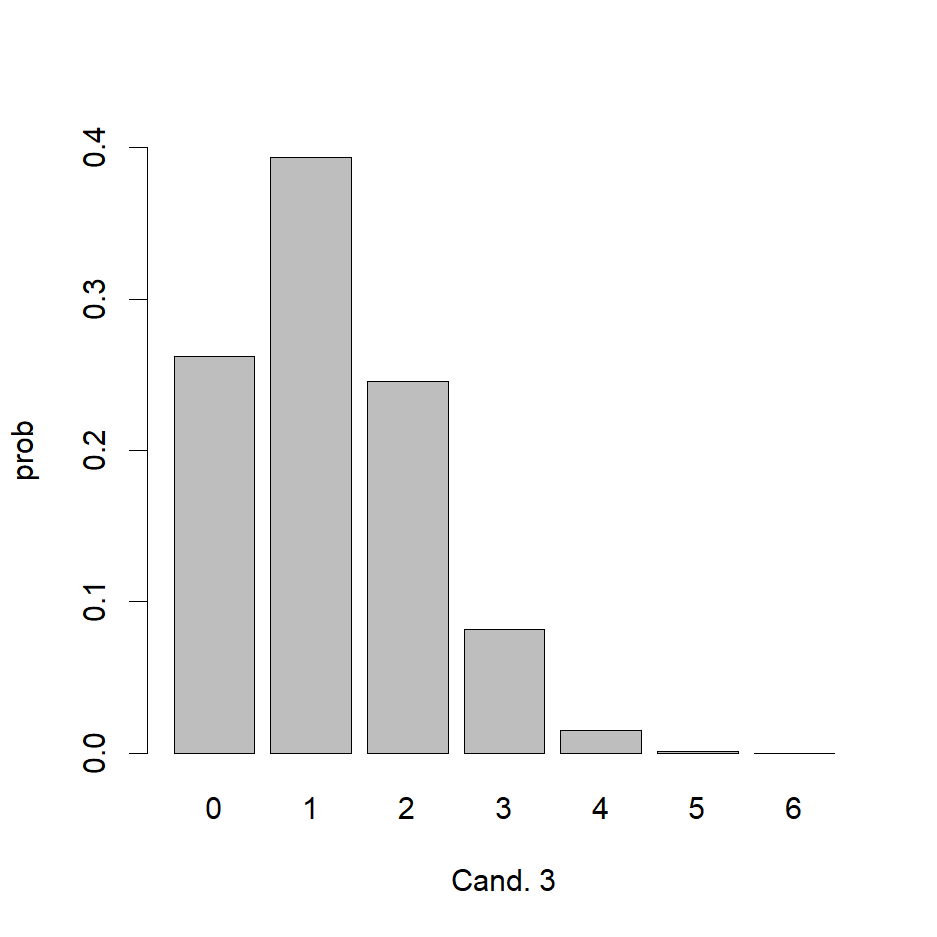}
    \caption{Barplots of evaluations for each of the  $m=3$ candidates  in a DDD multinomial model with $K=6$ and a probability vector (0.5, 0.3, 0.2).}

    \label{fig:exempleMultinomial}
   \end{figure}

Figure~\ref{fig:exempleMultinomial} shows an example of a DDD multinomial model for 3 candidates and $K=6$, with the probability vector (0.5, 0.3, 0.2). 

The continuous counterpart of the multinomial distribution is obtained through the use of a Dirichlet distribution on $[0,1]^m$  as follows (see \cite{dirichlet} for details about Dirichlet distribution).

 \begin{defn}{\bf DDD  Dirichlet model}~\\
\label{def:dirich}
The DDD  Dirichlet model is defined by
$(E_1, \ldots, E_m) \sim \mathcal{D}ir\{\alpha_1, \ldots, \alpha_m\}$, 
 where $\mathcal{D}ir\{\alpha_1, \ldots, \alpha_m\}$ is the Dirichlet distribution of parameters $\alpha_1,\ldots, \alpha_m$, with for all $c=1,\ldots,m$, $\alpha_c > 0$.
\end{defn}

The Dirichlet distribution is an extension of the Beta distribution to the multivariate case. The marginal distributions are Beta distributions and Dirichlet modeling introduces dependence between the evaluations of the candidates; the higher $\alpha_c$, the higher the mean score of candidate $c$, $c=1,\dots, m$.  If parameters $(\alpha_c)_{c=1,\ldots,m}$ do not have all the same value, then the Dirichlet model is an DDD model. If for all $c =1, \ldots, m$, $\alpha_c=1$, then the Dirichlet model is an DID model. We refer to \cite[Chapter 1 and Chapter 2]{dirichlet} and \cite{lin2016dirichlet} for an overview on Dirichlet distribution.

\begin{figure}[!ht]
    \centering
      \includegraphics[width=0.3\textwidth]{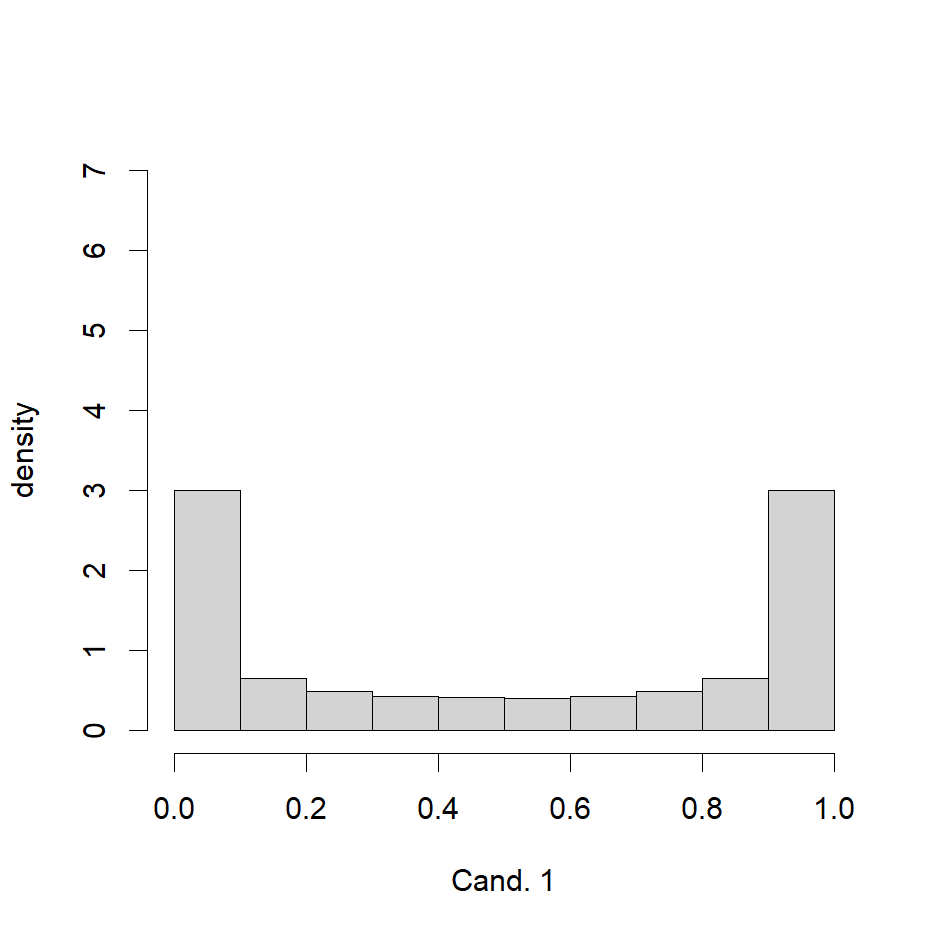}
    \includegraphics[width=0.3\textwidth]{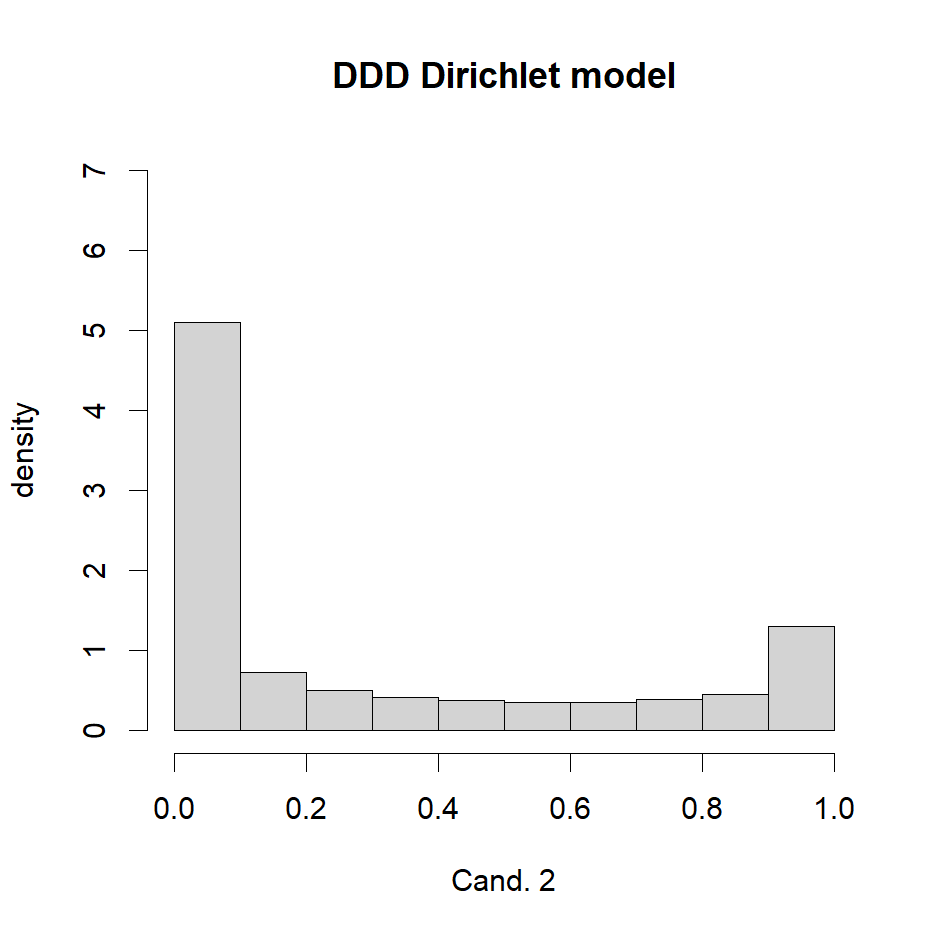}    \includegraphics[width=0.3\textwidth]{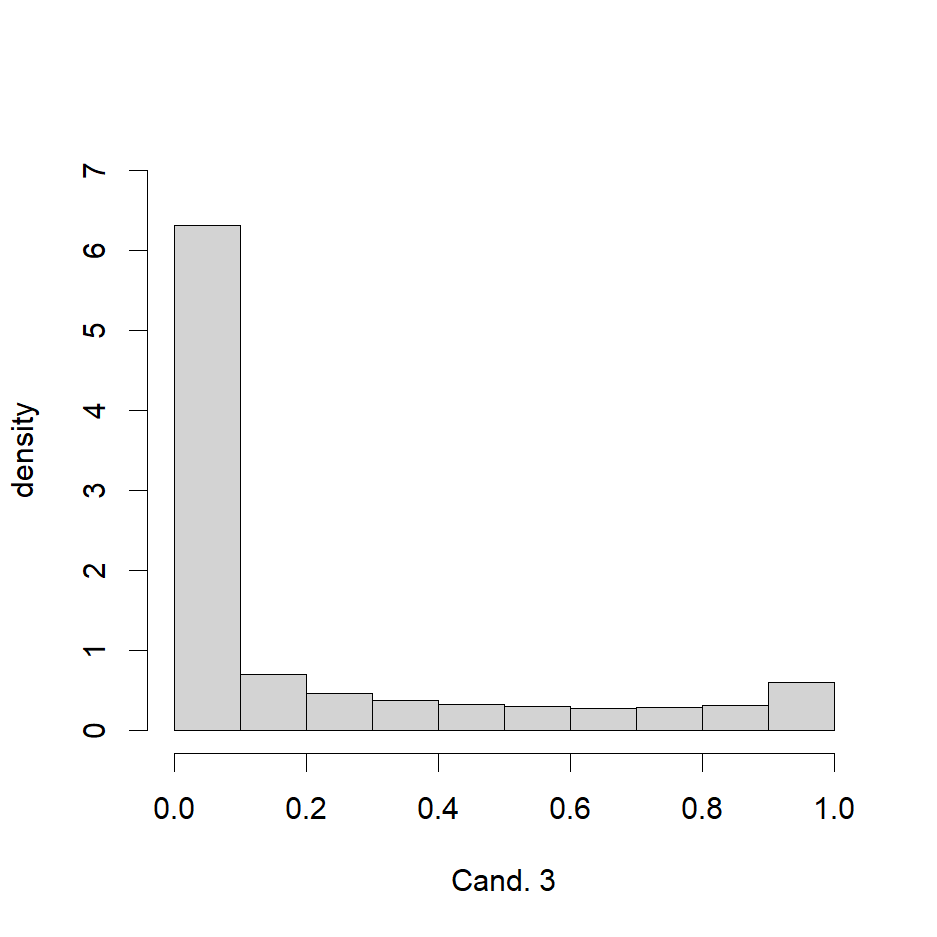}
        \caption{Histograms of evaluations for each of the $m=3$ candidates in a DDD Dirichlet model with parameters vector (5, 3, 2).}
    \label{fig:exempleDirichlet}
   \end{figure}

\begin{figure}[!ht]
    \centering
   \includegraphics[width=0.95\textwidth]{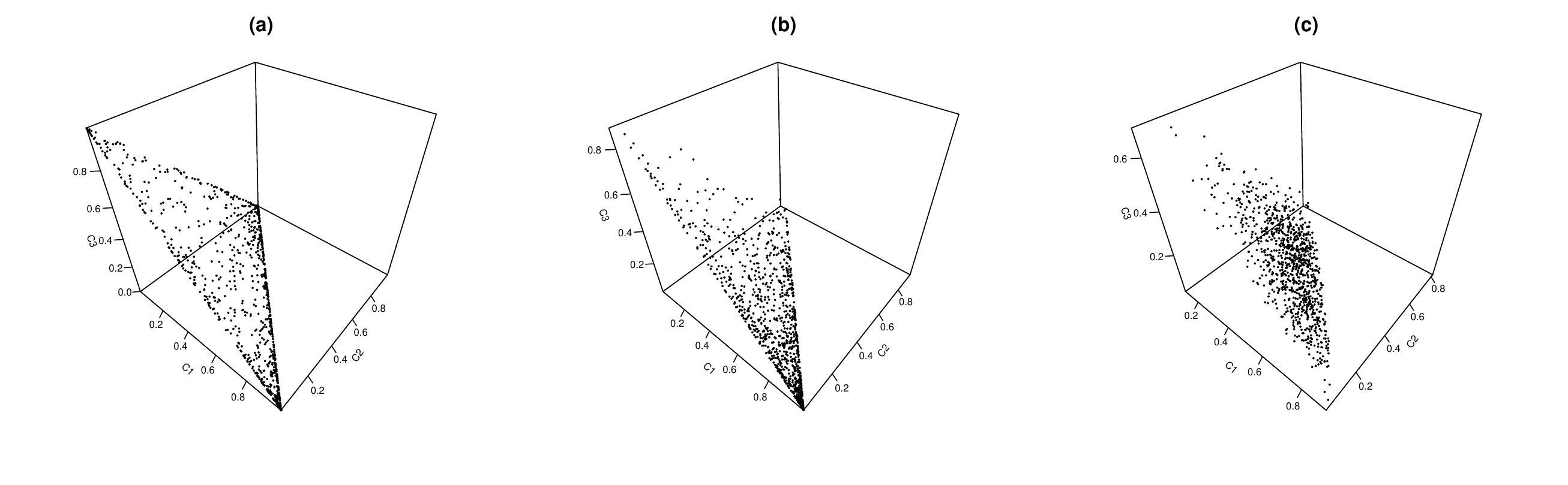}

\caption{Simulations of evaluations for each of the $m=3$ candidates ($n=1000$ voters) in a DDD Dirichlet model with (a) $(\alpha_1, \alpha_2, \alpha_3)=(0.5,0.3,0.2)$, (b) $(\alpha_1, \alpha_2, \alpha_3)= (2,0.5,0.5)$, (c) $(\alpha_1,\alpha_2,\alpha_3)=(5,3,2)$.}
    \label{fig:exempleDirichlet3d}
   \end{figure}

An example of simulation with 3 candidates using different probabilities for the candidates is shown in Figure~\ref{fig:exempleDirichlet}. The links between the  evaluations of the three candidates is shown in Figure~\ref{fig:exempleDirichlet3d}.
   
\subsection{Spatial models}\label{sec:spatial}

Spatial voting simulations have been developed following the early work of \cite{Downs1957} for votes based on rankings. The model is based on the use of an euclidean distance between the candidates and the voters, living in the same uni- or multi-dimensional space: the smaller the distance, the better the rank. \cite{Tideman2012b} conclude that a spatial model ``describes the observations in data sets much more accurately'' than other models.

We propose an adaptation of the spatial model in the framework of evaluation-based voting. Let $d$ be a given dimension parameter. Parameter $d$ should be seen as the number of latent characteristics which are used to build an opinion on the candidates. Typically, $d=2$ or $3$, see \cite{Armstrong2020}  for a discussion on the choice of $d$.  Voters 
and candidates 
are then randomly generated as points inside the hyperspace $\mathbb R^d$.

Spatial voting is next based on the distances between the generated points. The closer a voter to a candidate, the higher their evaluation of this candidate.

\begin{defn}{\bf DDD Spatial models}~\\ \label{def:spatial}
Let $x_v$,  $v=1,\dots,n$, and  $y_c$, $c=1,\dots,m$  be independent realizations from a distribution on $\mathbb R^d$,  $d\in\mathbb{N}\setminus \{0\}$.
The DDD spatial model for the evaluation $e_{vc}$ of candidate $c$ by voter $v$ is defined as 
$$  \forall v= 1,\ldots, n, \forall c=1,\ldots , m,\quad   e_{vc}=f(\delta(x_v,y_c)),$$
where $\delta$ is a distance between $x_v$ and $y_c$ and $f$ a non-increasing function mapping $\mathbb{R}^+$ to $[0,1]$.
\end{defn}

\begin{figure}[!ht]

\begin{multicols}{2}
    \centering

    \includegraphics[width=0.45\textwidth]{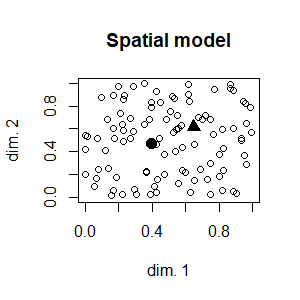}
        \caption{Simulation example of spatial model, with $m=2$ candidates ($\blacktriangle$ is candidate 1 and $\bullet$ is candidate 2) and $n=100$ voters in a 2-dimensional space ($d=2$), obtained through an Uniform distribution on $[0,1]^2$.}
    \label{fig:spatial1}

    \centering
    \includegraphics[width=0.45\textwidth]{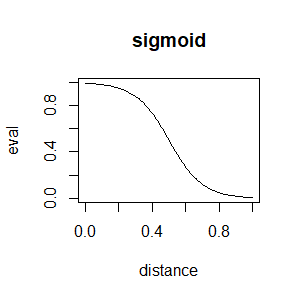}
        \caption{Example of sigmoïd transformation, $\delta \mapsto \bigl(1+e^{\lambda(\beta\delta-1)}\bigr)^{-1}$, with $\lambda=5$ and $\beta=2$.}
    \label{fig:sigmoid}
    
    \end{multicols}
\end{figure}

Typically, an intuitive spatial simulation model is given by the choices of \begin{itemize}
\item {a spatial distribution for the voters and the candidates}, $x_v$, $v=1,\dots,n$ and $y_c$, $c=1,\dots,m$. A classical choice is the Uniform distribution without any additive information about the voters, see Figure~\ref{fig:spatial1}. More specific distributions like a Gaussian distribution (resp. a mixing of Gaussian distributions) allow to obtain a bigger concentration of voters in a specific area of the hyperspace (resp. different areas), see e.g. \cite{SpatialNonUnif}.

\item {a distance}
A usual distance is the Euclidean distance $\delta_e$. Numerous distances exist, each with specific properties.

\item {a link function}, $f$. For instance, for $ v=1, \ldots, n$ and $c=1,\ldots , m$, $e_{vc}=\max\{0,(1-\ell \times \delta_e(x_v,y_c))\}$ with $\delta_e$ the Euclidean distance and $\ell > 0$. The parameter $\ell$ defines the decreasing rate of the evaluations with respect to the distance. For example, $\ell$ greater than $2$ ensures that a voter being anywhere on the frontier of the unit cube will give a null score to a candidate who is on the center of the unit cube or, equivalently, that a voter who is on the center of the cube will give a null score to an extreme candidate on the frontier of the cube. 

Other link functions $f$ are also possible, as for example the sigmoïd which is defined as follows:  for $ v=1,\ldots, n$, $c=1,\ldots, m$, $e_{vc}=\bigl(1+e^{\lambda(\beta \delta_e(x_v, y_c)-1)}\bigr)^{-1}$, with $\lambda>0$ and $\beta>0$. Figure~\ref{fig:sigmoid} presents an example of such a function for $\lambda=5$ and $\beta=2$. 
\end{itemize}

\begin{figure}[!ht]

    \centering

    \includegraphics[width=0.3\textwidth]{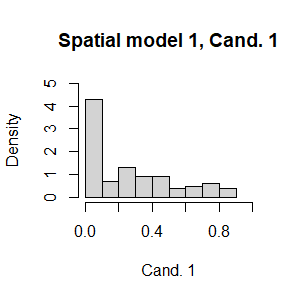}
    \includegraphics[width=0.3\textwidth]{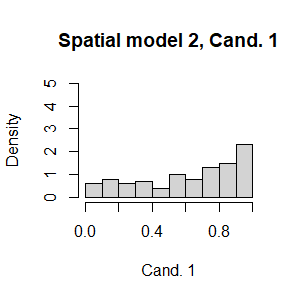}
    \includegraphics[width=0.3\textwidth]{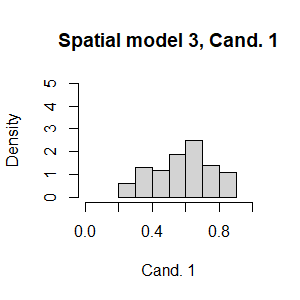}
    \includegraphics[width=0.3\textwidth]{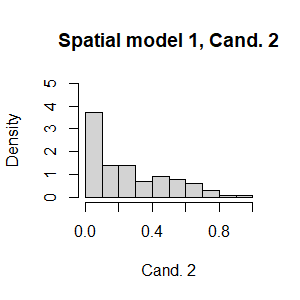}
    \includegraphics[width=0.3\textwidth]{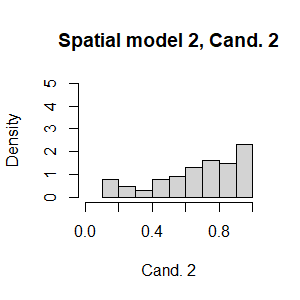}
    \includegraphics[width=0.3\textwidth]{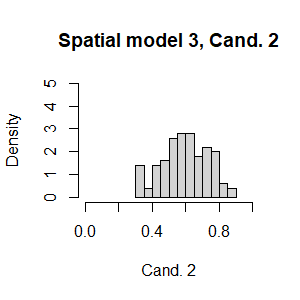}
        \caption{Histograms of evaluations obtained through the spatial model of Figure~\ref{fig:spatial1} and calculation models 1, 2 and 3 defined above.}
    \label{fig:spatial_eval}
\end{figure}

\begin{figure}[!ht]
\centering

    \includegraphics[width=0.3\textwidth]{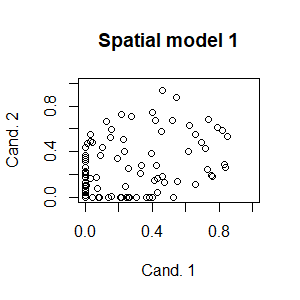}
    \includegraphics[width=0.3\textwidth]{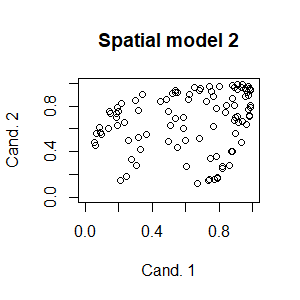}
    \includegraphics[width=0.3\textwidth]{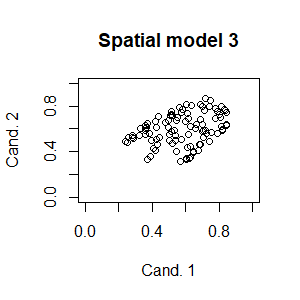}
        \caption{Plot of evaluations of candidate 2 vs candidate 1 given by each voter, with calculation models 1, 2 and 3 defined above.}
    \label{fig:spatialcorrelation}

\end{figure}

For a given position of candidates in $[0,1]^d$, the spatial model is clearly an DDD model, since the marginal distributions are different (see for example Figure~\ref{fig:spatial_eval}) and potentially correlated (see for example Figure~\ref{fig:spatialcorrelation}).

An example of evaluations obtained through a spatial model is presented in Figure~\ref{fig:spatial1}. Two candidates and $n=100$ voters have been randomly generated in $[0,1]^2$, that is, with $d=2$. Evaluations are then obtained from the euclidean distance of each voter to each candidate. We propose in Figure~\ref{fig:spatial_eval} histograms of three different ways to obtain evaluations, based on the same spatial situation (the one described in Figure~\ref{fig:spatial1}), for each candidate:
    \begin{itemize}
    \item evaluations in model 1 are obtained using $e_{vc}=max\{0,(1-2 \times \delta_e(x_v,y_c))\}$,
    \item evaluations in model 2 are obtained using $e_{vc}=(1+e^{\lambda(\beta \delta_e(x_v, y_c)-1)})^{-1}$ with $\lambda=5$ and $\beta=2$,
    \item evaluations in model 3 are obtained using $e_{vc}=(1+e^{\lambda(\beta \delta_e(x_v,y_c)-1)})^{-1}$ with $\lambda=2$ and $\beta=2$.
    \end{itemize}

Spatial models with discrete evaluations on $\{0,\dots, K\}$ can easily be obtained from continuous models by dividing the $[0,1]$ interval onto the $K+1$ intervals. Then if the continuous grade obtained with the spatial model belongs to the $l^\text{th}$ interval, the discrete evaluation is set to $l-1$. That is, for a continuous evaluation $e_{vc}$ on $[0,1]$, we consider $\lfloor (K+1)e_{vc}\rfloor$ as the resulting discrete evaluation, where $\lfloor u\rfloor$ denotes the greatest integer lower than $u$. The spatial interpretation of such a process is to determine $K$ spheres centered on the candidate and to give the evaluation of $K$ if the voter is into the smallest sphere, $K-1$ for the second smallest sphere and so on, until 0.

Note that the spatial model is interesting for generating evaluations since it offers various dependence structures between voters. Moreover, it is easily interpretable, with the latent space $\mathbb{R}^d$ seen as the space of the criteria used for discriminating the candidates. Considering the fitting power of spacial model, its non-parametric nature  makes it a very versatile tool, but with a high risk of overfitting. Yet, a drawback of this modeling is the fact that we cannot explicit the resulting distribution on evaluations, unlike when using parametric copulas models.

\section{Fitting real data}
\label{sec:geneobs}

We determine which models defined in the previous sections fit better real data sets.

In the easiest cases, a parametric approach will fit the real data in a satisfactory way. It will be very convenient for the interpretation of the study. For example, mixture models can be a solution to deal with heterogeneous situations and capture the complexity of real life situations. But a danger is then to overfit the data with too much parameters. A non parametric approach can also be considered. Yet, it can lead to more complicated interpretation of the model.

\begin{rem} Note that a Bayesian procedure could also be used to fit the models (see e.g. \cite{bayesian1,bayesian2,bayesian3} for Bayesian estimation of Thurstone order statistics models). We rather use a frequentist approach here. \end{rem}

Except for spatial modeling, a guideline for the choice of a specific model could be the following:
\begin{enumerate}
    \item Choose a general parametric model for each distribution and fit the marginal distributions of the evaluations of each candidate, $E_1,\ldots,E_m$ to the chosen model.
    \item Test if these marginal distributions can be considered as identical.
    \item Test the independence of $E_1, \ldots, E_m$ and propose a dependence modeling if the evaluations are not independent.
    \item Build the final model, using previous steps. 
\end{enumerate}
The results of the distributions equality and independence tests lead to choose the appropriate model within IID, IDD, DID and DDD models.

This guideline is relevant only in case of small amount of data, as it is well-known that statistical tests are inclined to systematically reject the null hypothesis when the sample size is too large, see \cite{Lantz, Mingfeng}. So we will not do any goodness-of-fit hypothesis test, but rather compare the quality of fitting for different models by providing a distance statistic between the observed distribution and the modeled one.

As a matter of illustration, we propose in the following two examples of fitting real voting situations through the proposed models. The first one deals with continuous evaluations, whereas the  other focus on discrete evaluations.

\subsection{Continuous case}

The first example, in a continuous framework, is based on the use of a survey concerning the 2017 presidential election in France. Data are available in \cite{bouveret_sylvain_2018_1199545}, and deal with 13 candidates evaluations on $\{0,\dots,100\}$  by $n=20210$ voters. 
We transform this 0-100 scale into a continuous scale, adding to any value given by a voter to a candidate a random value uniformly distributed between $0$ and $1$. These values included in the $[0,101]$ interval are then scaled to the $[0,1]$ interval.

\begin{figure}[!ht]
\centering
    \includegraphics[width=\textwidth]{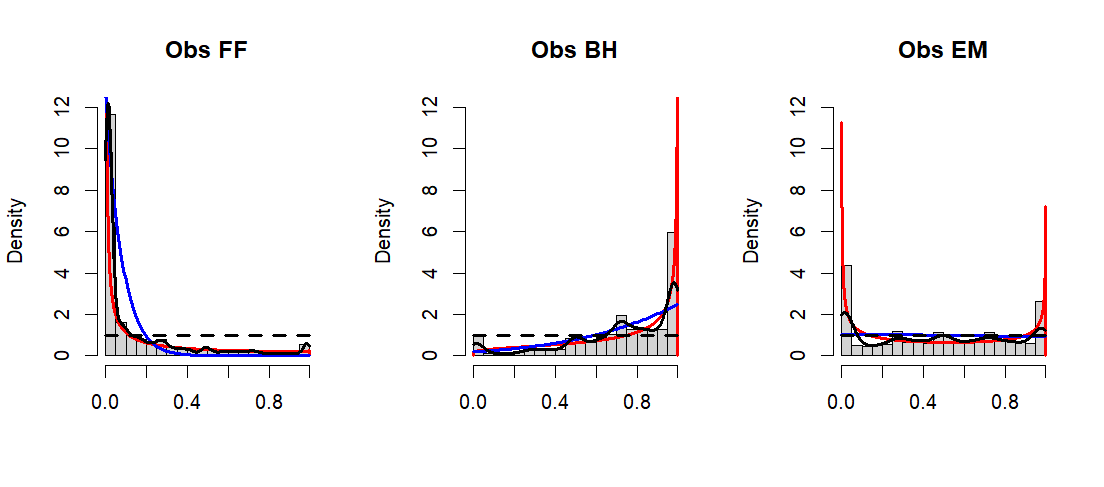}
        \caption{Histograms of observed evaluations of $m=3$ candidates given by $n=20210$ voters at the French 2017 election. Uniform model distribution is in dashed black, Beta model in red, truncated Normal in blue and checkerboard copula in solid black.}
    \label{fig: resultat_fit_continu} 
    \end{figure}

We focus on $m=3$ candidates: François Fillon (FF), Benoit Hamon (BH) and Emmanuel Macron (EM) (who finally won the election). Illustrations of the observed distributions are presented in Figure~\ref{fig: resultat_fit_continu}, with the distributions of Uniform, truncated Normal, Beta and checkerboard copula models.

\paragraph*{ Step 1. Fit a parametric model for each candidate's evaluations.}~\\

When choosing a general parametric model for each marginal evaluations distributions, considering empirical evaluations distributions is the first thing to do. The marginal distributions are simply the sample distributions, divided into $G$ classes. We tried $G=101$ to recover the initial discrete values on the 0-100 scale, and, as suggested in \cite{Cuberos2020}, we also tried $G=40$. In practice this approach is often efficient to fit a dataset but not appropriate to generate a predictive modeling, as the risk of overfitting is strong.

We then propose to model the marginal distributions using either Uniform distributions, Beta distributions or Truncated Normal distribution as a matter of example. Obviously, truncated Normal distribution does not seem accurate with these data, since, as illustrated in Figure~\ref{fig: resultat_fit_continu}, the shapes of the distributions are far from a Gaussian curve.

The Uniform distribution does not need any parameter estimation. The Beta distribution  depends on two parameters $\alpha$ and $\beta$. These parameters can be estimated using the method of moments. Let $\hat{\mu}$ denote the sample mean and $s$ the sample standard deviation. Then parameters $\alpha$ and $\beta$ can be estimated respectively by $\hat\alpha =\hat{\mu}\left({\frac {\hat{\mu}(1-\hat{\mu})}{s^2}}-1\right)$ and $\hat \beta =(1-\hat{\mu})\left({\frac {\hat{\mu}(1-\hat{\mu})}{s^2}}-1\right)$.
The truncated Normal distribution depends on two parameters, the mean and standard deviation, which are estimated via a maximum likelihood method (see \cite{CRAIN}), using the package developed by \cite{tmvtnorm}.

The first three lines of Table~\ref{table:resultat_fit_continu} gives the Kolmogorov-Smirnov statistic, corresponding to the distance between the distribution of observed evaluations and the distribution of simulated evaluations. The smaller the distance, the more the model fits the sample distribution. 
Concerning the non parametric approach, the statistics are presented in the last two lines of Table~\ref{table:resultat_fit_continu}. It is obvious that the empirical distribution with $G=101$ classes overfits the data. The empirical distribution with $G=40$ classes is competitive with the Beta distribution.

\begin{table}[!ht]
\centering
\begin{tabular}{rd{1.3}d{1.3}d{1.3}}  \hline
 & FF & BH & EM \\ \hline
Uniform marginals dist. & 0.570 & 0.354 & 0.180 \\
Trunc. normal marginals dist. & 0.492  & 0.344 & 0.179 \\ 
Beta marginals dist. & 0.269  & 0.099 & 0.119 \\
\hline
Sample dist. divided into 40 classes & 0.267  & 0.142 & 0.105 \\
Sample dist. divided into 101 classes & 0.006  & 0.007 & 0.005 \\ \hline
\end{tabular}
\caption{Kolmogorov-Smirnov statistic for 5 fitting models. \label{table:resultat_fit_continu}}
\end{table}

\paragraph*{Step 2. Test if these marginal distributions can be considered as identical.}~\\

It is straightforward from Figure~\ref{fig: resultat_fit_continu} that the marginal distributions are different, and, hence, that the modeling of the data is either DID or DDD. Note that in the general case, such test can be done with a Kruskall-Wallis test.

\paragraph*{Step 3. Test the independence of $E_1, \ldots, E_m$ and model the dependence if necessary.}~\\

Let us now focus on the independence between the evaluations. As one can see in Table~\ref{tab:pres_correl}, evaluations given to FF and BH are negatively correlated, whereas evaluations given to FF and EM are slightly positively correlated, and evaluations given to BH and EM are not correlated. Based on the observed correlation values, the independence assumption is not realistic. Formally, a Bartlett's sphericity test on the correlation matrix can be applied to check the dependence property. We, hence, consider a DDD model. We consider below, as a matter of example, copulas models to capture the dependencies between the evaluations given to the candidates.

\begin{figure}[!ht]
\centering
    \includegraphics[width=0.33\textwidth]{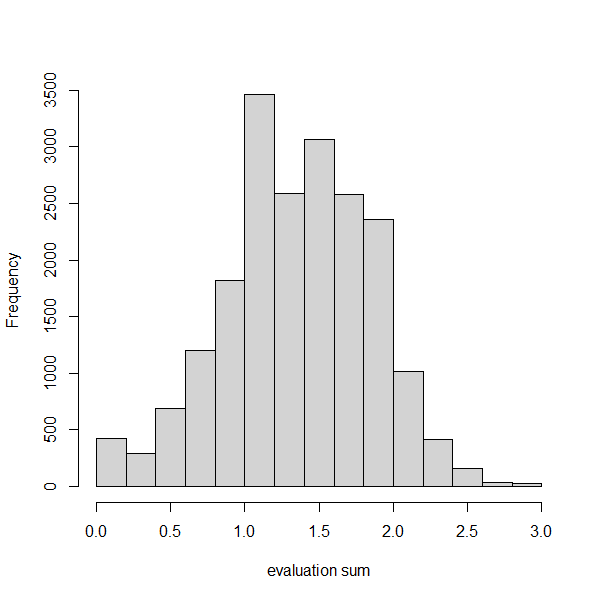}
        \caption{Histogram of the sum of the evaluations given by each voter to the candidates}
    \label{fig:histsum} 
    \end{figure}

\begin{rem} 
Dirichlet model deals with cumulative voting, where the evaluations provided by each voter for all candidates collectively sum to a constant. Obviously this is not the case here, as shown in Figure~\ref{fig:histsum}. 
\end{rem}

We consider two Copula-based models.
\begin{itemize}
    \item {\bf Parametric Copula model.} We consider a Gaussian copula, with marginal distributions fitted by Beta distributions as established above, and correlation coefficients equal to the observed correlation coefficients shown in Table \ref{tab:pres_correl}. It has the advantage to be parametric and to provide a reproducible modeling.
  \item {\bf Non parametric Copula model.}  We consider a checkerboard Copula model, which does not require any assumption on the marginal distributions. The checkerboard copula model is simply the multivariate sample distribution, divided into $B^m$ classes, where the marginal distributions are determined as above. The value of $B$ in the following is 40. Note that $B=101$ leads to the empirical copula which overfits the data.
  \end{itemize}

Non-parametric Copula model exhibits a superior fit to the data compared to the parametric Copula model, particularly when the number of classes is sufficiently large. This outcome is unsurprising, given the inherent parametric/non-parametric nature of the respective models.

Table \ref{tab:pres_correl_copula} displays the correlation coefficients obtained with the two modelings.

\begin{table}[!ht]
\centering
\begin{tabular}{ld{1.4}d{1.4}d{1.4}} \hline
        &   FF    &        BH  &         EM \\ \hline
FF &  1 &-0.41 &0.30 \\
BH & -0.41 &  1 & 0.0008 \\
EM &  0.30 & 0.0008 & 1 \\ \hline
\end{tabular}
\caption{Empirical correlations between three candidates at the French 2017 presidential election.}
\label{tab:pres_correl}
\end{table}

\begin{table}[!ht]
\centering
\begin{tabular}{ld{1.3}d{1.3}d{1.3}} \hline
        &   FF    &        BH  &         EM \\ \hline
FF &  1 &-0.33 & 0.23 \\
BH & -0.33 &  1 & 0.002 \\
EM &  0.23 & 0.002 & 1 \\ \hline
\end{tabular}
\hspace{1cm}
\begin{tabular}{ld{1.2}d{1.2}d{1.2}} \hline
        &   FF    &        BH  &         EM \\ \hline
FF &  1 &-0.41 &0.30 \\
BH & -0.41 &  1 & 0.01\\
EM &  0.30 & 0.01 & 1 \\ \hline
\end{tabular}
\caption{Correlations between candidates obtained with copula DDD models at the French 2017 presidential election. On the left, correlations obtained with a Normal copula and Beta marginals, on the right correlations obtained through the use of a checkerboard copula with $B=40$ classes associated with empirical marginal distributions with $G=40$ classes.}
\label{tab:pres_correl_copula}
\end{table}

\paragraph*{Step 4. Build the final model, using previous steps.}~\\
As a conclusion, in this case Copula models seems more appropriate with Beta marginal distributions. A non-parametric approach of course better fits the data distributions, but a parametric approach seems more appropriate to model the situation without overfitting.

\paragraph*{Spatial representation}

The spatial model introduced in Section \ref{sec:spatial} can also be used as a representation of candidates and voters in the same space. However, the experiments show that it is difficult to simulate new data from a spatial representation: even if the spatial representation of the voting situation is accurate, it is not an easy task to identify the latent space (its dimension $d$ and its metric) nor the distribution of the voters in the voting space. Suppose the dimension $d$ known, that the metric is given, and let us focus on the distributions of voters in the latent space. We propose a two steps process, to simulate new data:
\begin{enumerate}
\item estimate both candidates and voters positions into a $d$-dimensions space, for example by the use of the SMACOF method, see \cite{smacof},
\item estimate the distribution of voters into the $d$-dimensions space, in order to generate new voters with the same distribution.
\end{enumerate}
The second step needs to fit a multidimensional distribution. The fitting is not done directly on the data but on the latent positions obtained with the first step. This can be done for example with a Copula model, as previously.

\begin{figure}[!ht]
\centering
     \includegraphics[width=0.32 \textwidth]{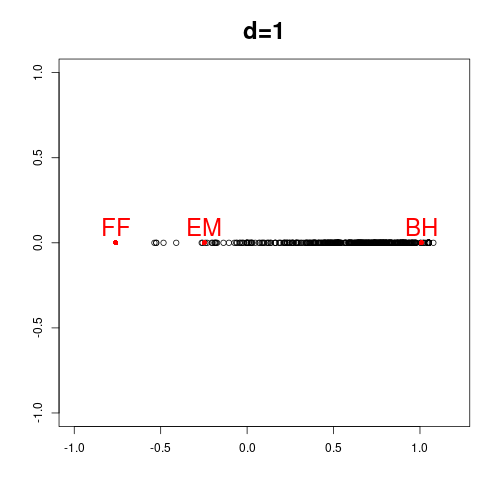}
    \includegraphics[width=0.32 \textwidth]{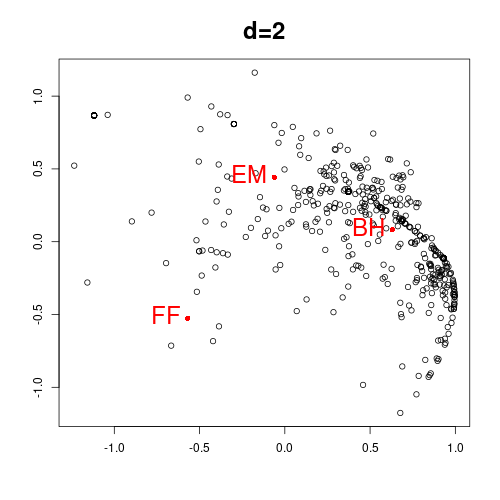}
    \includegraphics[width=0.32 \textwidth]{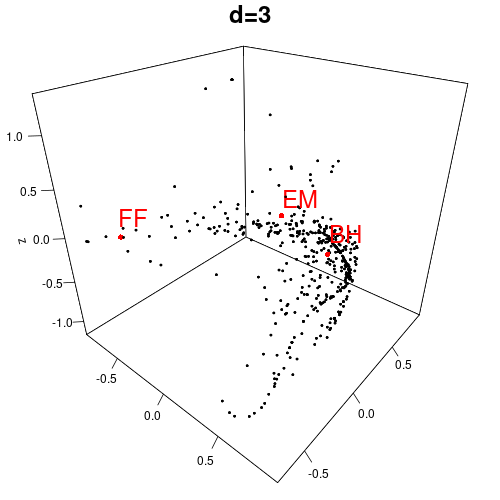}
    \caption{Spatial representation of candidates (in red) and voters of the French 2017 election with by the SMACOF algorithm, in dimension $d=1$, $d=2$ and $d=3$, from left to right.
    }
    \label{fig:spatial_elections} \end{figure}

We choose hereafter to consider $d=1$, $d=2$ and $d=3$, and the euclidean distance. We refer to \cite{dimension} for the choice of $d$. As stated therein, the choice is not easy. The higher $d$, the higher the complexity of the model. Actually, choosing $d\leq 3$ allows an easy visualization of the latent space. The spatial representations obtained by the SMACOF algorithm are shown in Figure~\ref{fig:spatial_elections}. As shown by the figure, the distribution with $d=3$ does not differ much to the distribution obtained with $d=2$. Considering $d=2$ seems sufficient to capture the information on the evaluations. This supports the conclusion of \cite{Armstrong2020} that small values of $d$ are often sufficient. 

Hence, from real data, we have calibrated the parameters of a spatial model, which allows generating new data. Note that the intrinsic dimension and the used distance are to be set by the user. Some post-hoc measures of the quality of the adjustment are available for comparing different choices, see \cite[chapter 4]{smacof}.

\subsection{Discrete case}

We study in this section voting data collected in the general assembly of a french scout association where 7 options have been presented to 93 voters for a specific resolution\footnote{Data are available in the supplementary material. The association is the \textit{Eclaireuses and Eclaireurs Unionistes de France} and the general assembly stood in January 2022 in Bordeaux, France. Data are from private communication. 
}.
 The voting procedure used in this assembly was the majority judgment with evaluations on a 7 levels discrete scale.

\paragraph*{ Step 1. Fit a parametric model for each candidate's evaluations.}~\\

We first propose to model the marginal distributions on the candidates using Uniform distributions, Binomial distributions and Beta-Binomial distributions.

The Binomial distribution needs the estimation of a parameter $p$, which can be estimated by $\hat{p}=\frac{\hat{\mu}}{K+1}$, where $\hat{\mu}$ is the sample mean and $K+1$ is the number of scales  in the evaluations. The Beta-Binomial distribution needs the estimation of parameters $\alpha$ and $\beta$, obtained as follows:  $\hat\alpha=\frac{K\hat{\mu}-\hat{\mu}^2-s^2}{K(s^2/\hat{\mu}-1)+\hat{\mu}}$ and $\hat\beta=\frac{(K-\hat{\mu})(K-\hat{\mu}-s^2/\hat{\mu})}{K(s^2/\hat{\mu}-1)+\hat{\mu}}$, with $\hat{\mu}$ the sample mean and $s$ the sample standard deviation, see \cite{beta-binomial-estimation}.

For each candidate, we compute the $\chi^2$ statistics of the distances between the distribution of observed evaluations and the theoretical Uniform, Binomial and Beta-Binomial distributions. The smaller the distance, the better the model fits the observed distribution.

\begin{figure}[!ht]
\centering
    \includegraphics[width=\textwidth]{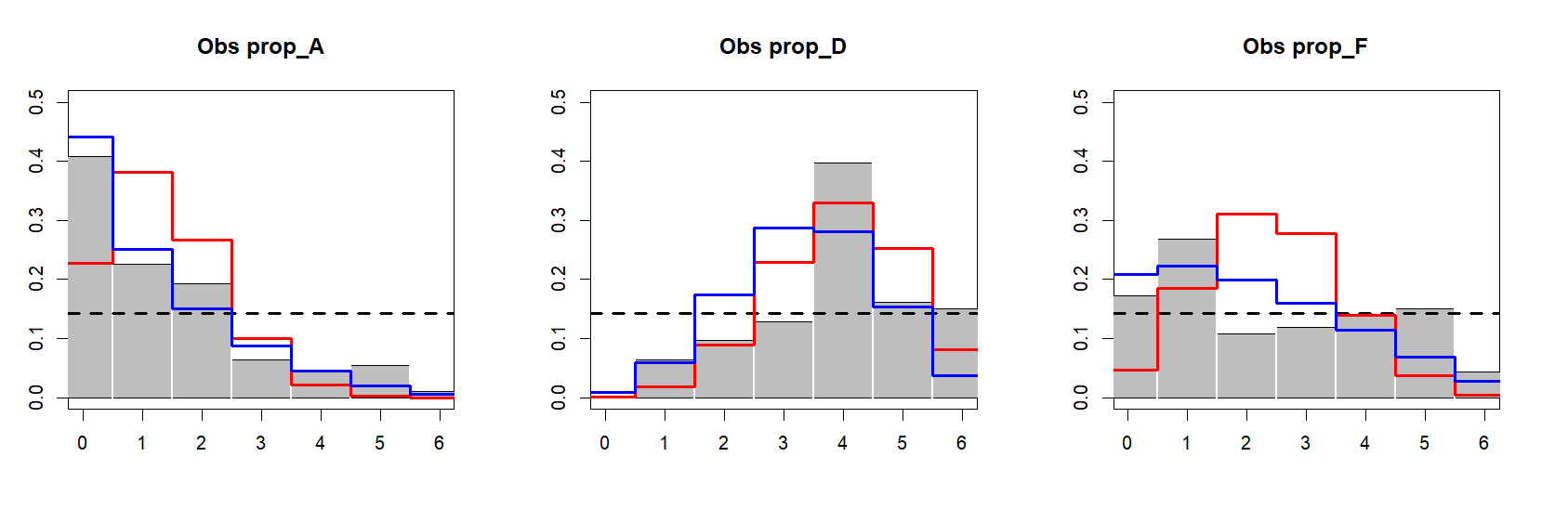}
        \caption{Bar plots of observed values   for  propositions C, F, and G. The dashed black line represents the uniform distribution. The distribution in red corresponds to the binomial model, and the distribution in blue corresponds to the beta-binomial model.}
    \label{fig:resultat_fit_discret} 
\end{figure}

Table~\ref{table:resultat_fit_khi2} shows the summary statistics of the  $\chi^2$ distances between the observed distributions of the evaluations and the simulated distributions for the 7 candidates. The Beta-Binomial distribution has a better fit than the Binomial or the Uniform distributions. Bar plots for observed distributions and models for three propositions are displayed in Figure~\ref{fig:resultat_fit_discret}.

\begin{table}[!ht]
\centering
\begin{tabular}{ld{3.2}d{2.2}d{2.2}d{2.2}d{2.2}d{2.2}d{4.2}} \hline
 Proposition &  A & B & C & D  & E  & F   & G\\ \hline 
binomial & 225.98 & 58.11 & 41.19 & 24.81 & 163.46 & 121.74&  2042.11 \\
uniform   & 79.14 & 16.82 & 23.90 & 61.38 &  21.33 &  18.62 &  100.21 \\
beta-binomial &  8.23 &  18.91&  18.30&  48.94 &  27.10 &  16.88 &    9.60 \\ \hline
\end{tabular}
\caption{Summary of $\chi^2$ distances between the observed distributions of the evaluations and the simulated distributions for the 7 candidates.}
\label{table:resultat_fit_khi2}
\end{table}

\paragraph*{Step 2. Test if these marginal distributions can be considered as identical.}~\\

Figure~\ref{fig:resultat_fit_discret} shows that the hypothesis of identical distributions cannot be assumed. We will hence consider DID or DDD models.

\paragraph*{Step 3. Test the independence of $E_1, \ldots, E_m$ and model the dependence if necessary.}~\\

On the 21 pairs of candidates 9 have a negative correlation, and 12 have a positive correlation. The correlation coefficient is significantly different to zero for 7 pairs of candidates. This shows the importance of taking into account dependence in modeling. Therefore, as in the continuous case, the evaluations in the discrete case are not independent and we will consider a DDD model. 

\begin{figure}[!ht]
\centering
    \includegraphics[width=\textwidth]{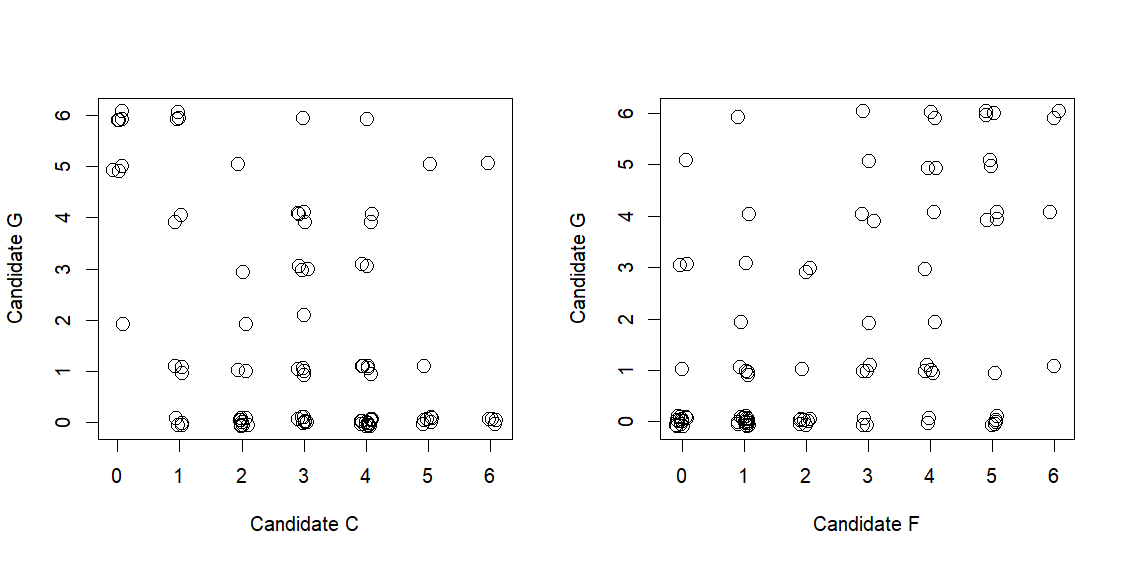}
\caption{Correlation plots of observed evaluations between candidates C and G (left) and candidates F and G (right).}
    \label{fig:3D_correl_eeudf} 
    \end{figure}

As a matter of example, Table~\ref{tab:eeudf_correl} shows the correlations between the three candidates C, F and G. 
First, the marginal distributions are fitted as described above by a Beta-Binomial distribution. Then, a Gaussian discrete copula is used to model the correlation between candidates. A checkerboard copula based on $B=40$ classes has also been used, considering empirical distributions on the same $G=40$ classes of the copula. The obtained correlations are presented in Table~\ref{tab:eeudf_correl}, and bar plots are presented in Figure~\ref{fig:3D_correl_eeudf}.

\paragraph*{Step 4. Build the final model, using previous steps.}~\\
The discrete copula captures the dependence structure, but may introduce overfitting of the dependence, when generating simulated observations. No parametric modeling, among the ones proposed, are well adapted in such context. In this case, a non parametric approach seems more convenient.

\begin{table}[!ht]
\centering
\begin{tabular}{ld{1.2}d{1.2}d{1.2}} \hline
        &   C   &       F  &       G\\ \hline
C &  1 &-0.18 & -0.37 \\
F & -0.18 &  1 & 0.47\\
G &  -0.37 & 0.48 & 1 \\ \hline
\end{tabular}
\hspace{1cm}
\begin{tabular}{ld{1.2}d{1.2}d{1.2}} \hline
        &   C   &       F  &       G\\ \hline
C &  1 &-0.15 & -0.18 \\
F & -0.15 &  1 & 0.48\\
G &  -0.18 & 0.48 & 1 \\ \hline
\end{tabular}
\caption{Spearmann correlations between the evaluations of three candidates. On the left the observed correlations, on the right the correlations obtained with a Copula DDD Beta-Binomial model.}
\label{tab:eeudf_correl}
\end{table}

\begin{rem}
Note that a multinomial model is not accurate in the observed situation as the sum of scores for each voters is neither fixed nor equal to the maximum score. 
\end{rem}

\section{Conclusion}

As explained in the preamble, simulations can be done in two different settings.   
\begin{itemize}
    \item On the one hand, simulations can be done without any specific context, and the tuning of the distribution of the evaluations is let free or determined by external considerations. One has therefore to choose a model and set the parameters to arbitrary values. Examples of such simulation settings have been proposed with the description of each model above.
    \item On the other hand, one can wish to simulate observations in harmony with real data. In that case, an adjustment of the model to the observed data is necessary. 
     The aim is therefore 1) to choose the appropriate model 2) to infer the model parameters from the available data. This situation is detailed in Section \ref{sec:geneobs} for both discrete and continuous cases. 
\end{itemize}

We introduced in this paper several models to simulate evaluation-based voting data.
 Three main families of distributions were proposed for the marginal distributions of the evaluations, in a continuous setting and in a discrete setting. On the contrary to preference rankings models, where the key notion is the impartiality, a more refined discussion is needed for evaluation-based processes. 

The simplest model is when evaluations are independent and identically distributed (IID models). We propose first to distinguish either the marginal distributions are identical or not (IDD models). Such models do not imply Impartial Culture on preferences. Next, introducing dependence (DDD models) creates more complex models. We give examples of dependent distributions with identical marginals (DID models) which provide Impartial Culture on preferences. In particular, we introduce Copula DID and DDD models which allow to model the dependence between the evaluations. The variety of modeling described here offers the possibility of studying the properties of evaluation-based voting processes with an extensive probabilistic approach. It also provides new IC and non IC simulation approaches for preferences, since preferences rankings can be deduced from evaluations. Finally, as some proposed settings are parametric, the studied models can be fitted to real dataset to deduce more realistic frameworks. We present examples of such an approach on real data for continuous and discrete evaluations.

Finally, the \texttt{R} package \texttt{VoteSim} in \cite{voteSim}, available on the CRAN repository, aims at providing the models introduced in this paper for simulating distributions of evaluations given by voters on candidates concerning a voting situation within an evaluation-based perspective. The proposed simulation functions cover Uniform, Truncated Normal, Beta, Dirichlet, multinomial, copula-based and spatial models. Basically a function is proposed for each model which produces a matrix with candidates in rows and voters in columns. 

\bibliographystyle{plain}
\bibliography{Bibliosimu}

\end{document}